\def\rfr#1{Eq. (\ref{#1})}

\def\yu{\exp\left(-\rp{r}{\lambda}\right)}

\def\bar{\begin{eqnarray}}
\def\ear{\end{eqnarray}}

\def\eqi{\begin{equation}}
\def\eqf{\end{equation}}
\def\eqia{\begin{eqnarray}}
\def\eqfa{\end{eqnarray}}
\def\rp#1#2{{#1\over#2}}
\def\ct#1{\cite{#1}}
\def\lb#1{\label{#1}}
\def\pkp{ P^{(0)} }

\def\psr{PSR J0737-3039A/B\ }

\def\oc2{$\mathcal{O}(c^{-2})$}

\def\bds#1{\boldsymbol{#1}}

\documentclass{ws-ijmpd}

\begin{document}

\title{CONSTRAINING MODELS OF MODIFIED GRAVITY WITH THE DOUBLE PULSAR PSR J0737-3039A/B SYSTEM}

\author{LORENZO IORIO}
\address{Viale Unit$\grave{a}$ di Italia 68, 70125, Bari (BA), Italy
\\e-mail: lorenzo.iorio@libero.it}

 \author{MATTEO LUCA RUGGIERO}
  \address{Dipartimento di Fisica, Politecnico di Torino, Corso Duca degli Abruzzi 23, Torino, Italy\\
INFN, Sezione di Torino, Via Pietro Giuria 1, Torino, Italy\\
e-mail: matteo.ruggiero@polito.it}

\maketitle



\begin{abstract}
In this paper we use  $\Delta P = -1.772341\pm 13.153788$ s
between the phenomenologically determined orbital period $P_{\rm b}$ of the \psr\ double pulsar system and the
purely Keplerian period $P^{(0)}=2\pi\sqrt{a^3/G(m_{\rm A}+m_{\rm B})}$ calculated with the system's parameters,
determined independently of the third Kepler law itself, in order to put constraints
on some models of modified gravity ($f(R)$, Yukawa-like fifth force,
MOND). The major source of error affecting $\Delta P$ is not the one in the phenomenologically measured period
($\delta P_{\rm b}=4\times 10^{-6}$ s), but the systematic uncertainty $\delta\pkp$ in the computed Keplerian
one due to the relative semimajor axis $a$ mainly caused, in turn, by the errors in the ratio $\mathcal{R}$ of
the pulsars' masses and in $\sin i$. We get
$|\kappa|\leq 0.8 \times 10^{-26}$ m$^{-2}$ for the parameter that in the $f(R)$ framework is a measure of the
non linearity of the theory, $|\alpha|\leq 5.5\times 10^{-4}$ for the fifth-force strength parameter (for
$\lambda\approx a=0.006$ AU). The effects predicted by the strong-acceleration regime of MOND are far too small to be
constrained with some effectiveness today and in the future as well. In view of the continuous timing of such an
important system, it might happen that in the near future it will be possible to  obtain somewhat tighter
constraints.

\end{abstract}




\section{Introduction}
One of the six Keplerian orbital elements in terms of which it is
possible to parameterize the orbital motion of a pulsar in a
binary system is the mean anomaly  $\mathcal{M}$ defined as
$\mathcal{M}\equiv n(t-T_0)$, where $n$ is the mean motion and
$T_0$ is the time of periastron passage. The mean motion $n\equiv
2\pi/ P_{\rm b}$ is inversely proportional to the time elapsed
between two consecutive crossings of the periastron, i.e. the
anomalistic period $P_{\rm b}$. In Newtonian mechanics, for two
point-like bodies, $n$ reduces to the usual Keplerian expression
$n^{(0)}=\sqrt{GM/a^3}$, where $a$ is the semi-major axis of the
relative orbit of the pulsar with respect to the companion and
$M\equiv m_{\rm p}+m_{\rm c}$ is the sum of their masses. In
pulsar timing the period $P_{\rm b }$ is very accurately
determined in a phenomenological, model-independent way, so that
it accounts for all the dynamical features of the system, not only
those coming from the Newtonian point-like terms, within the
measurement precision.

In this paper we wish to look for phenomenologically determined deviations from the third Kepler law in the \psr system\cite{Bur03},
consisting of two radio-pulsars moving along a moderately eccentric,
2.4 hr orbit (see Table \ref{tavola} for its relevant orbital parameters),
in order to dynamically constrain some models of modified gravity  ($f(R)$, Yukawa-like fifth force, MOND).
\begin{table}[ph]
\tbl{
Relevant orbital parameters of the
\psr system$^2$.
The projected semimajor axis is
defined as $x=(a_{\rm bc}/c)\sin i,$  where $a_{\rm bc}$ is the
barycentric semimajor axis, the phenomenologically determined post-Keplerian parameter $s$
is equal to $\sin i$ in general relativity, and  $i$ is the angle
between the plane of the sky, perpendicular to the line-of-sight,
and the orbital plane. We  have
conservatively quoted the largest error in $s$ reported
in Ref.~2.
The parameter $e$ is the eccentricity. The orbital
period is known with a precision of $4\times 10^{-6}$ s. The
quoted values of $m_{\rm A}$ and $m_{\rm B}$ were obtained
in Ref.~2
by using the general relativistic expression of the
post-Keplerian A's periastron advance for the sum of the masses
$M=2.58708(16)$, and the ratio of the masses $\mathcal{R}\equiv m_{\rm
A}/m_{\rm B}=1.0714(11)$ phenomenologically determined from both
the projected semimajor axes $x_{\rm A}$ and $x_{\rm B}$ . The
rotational period ${\mathcal{P}_{\rm A}}=2\pi/\Omega_{\rm A}$ of
\psr A amounts to 22 ms, while ${\mathcal{P}_{\rm
B}}=2\pi/\Omega_{\rm B}=2.75$ s.
 }
{\begin{tabular}{@{}cccccc @{}} \toprule

$P_{\rm b}$ (d)& $x_{\rm A}$ (s) & $s$ & $e$ & $m_{\rm A}$ (M$_{\odot}$) & $m_{\rm B}$ (M$_{\odot}$)\\ \colrule
$0.10225156248(5)$ & $1.415032(1)$ & $0.99974(39)$ & $0.0877775(9)$ & $1.3381(7)$ & $1.2489(7)$\\
\botrule
\end{tabular}\label{tavola}}

\end{table}
We believe that the analysis presented here has the merit of quantitatively determining how useful a laboratory  such as \psr and a method like the orbital period can be in testing alternative theories of gravity  by precisely individuating the major sources of errors. Indeed, as many different laboratories and methods as possible should be used to extensively tests a fundamental interaction as gravity.
Such an analysis may also yield insights concerning future obtainable improvements in view of the continuous timing of such a binary system.
\section{Deviations from the third Kepler law}
Looking for genuine deviations from the third Kepler law$-$meant as discrepancy $\Delta P$ between the phenomenologically determined
orbital period $P_{\rm b}$ and the purely Keplerian one
$\pkp$ calculated with the values of Table \ref{tavola}$-$from the \psr timing data is possible because
all the parameters entering $\Delta P$ have been just
 measured independently of the third Kepler law itself. Indeed, the relative semimajor axis entering $\pkp$ \eqi a=\left(1+\mathcal{R}\right)\left(\rp{cx_{\rm A}}{s}\right)\lb{ohe}\eqf
is built in terms of the
ratio $\mathcal{R}$ of the masses, the projected semimajor axis $x_{\rm A}$ and $\sin i$; the projected semimajor
axes $x_{\rm A}$ and $x_{\rm B}$ were phenomenologically determined from the timing data, the phenomenologically estimated
post-Keplerian parameter $s$ can be identified with $\sin i$ in general relativity, the
ratio $\mathcal{R}$ has been phenomenologically determined from
the ratio of the projected semimajor axes coming from the quite
general relation, valid to at least the first post-Newtonian order\cite{Dam88,Dam92} \eqi a_{\rm bc}^{\rm (A)}m_{\rm A}=a_{\rm
bc}^{\rm (B)}m_{\rm B},\eqf and the sum of the masses
has been, in turn, derived from the A's periastron advance which is, at
present, the best determined post-Keplerian parameter. The third
Kepler law only enters the expression of the mass functions \eqi
{\mathcal{F}}_{\rm A/B}=\rp{m^3_{\rm B/A}\sin^3 i}{(m_{\rm
A}+m_{\rm B})^2}, \eqf which, instead, have not been used in
obtaining the parameters of \psr.

Table \ref{tavola} yields
\eqi
\Delta P\equiv P_{\rm b}-P^{(0)} = -1.772341\pm 13.153788\ {\rm
s}\lb{discr}.\eqf In \rfr{discr} we have evaluated
\eqi\delta(\Delta P)\leq \delta P_{\rm b}+\delta P^{(0)},\eqf with
\eqi\delta P^{(0)}\leq \delta P^{(0)}|_{a} + \delta P^{(0)}|_{M} +
\delta P^{(0)}|_{G}. \lb{errper}\eqf
The terms in \rfr{errper} are
\begin{equation}\left\{\begin{array}{lll}
\delta P^{(0)}|_{a} \leq 3\pi\sqrt{\rp{a}{GM}}\delta a=12.218572\ {\rm s}, \\\\
\delta P^{(0)}|_{M} \leq =\pi\sqrt{\rp{a^3}{G M^3}}\delta
M=0.273248\ {\rm s}, \\\\
\delta P^{(0)}|_{G} \leq =\pi\sqrt{\rp{a^3}{G^3 M}}\delta G=
0.661964\ {\rm s},
 \lb{katzi}
\end{array}\right.\end{equation}
so that
\eqi\delta\pkp\leq 13.153784\ {\rm s}.\eqf
In \rfr{katzi} we used the values of
Table \ref{tavola} for $\delta M$ and\cite{Moh05} $\delta G=0.0010\times
10^{-11}$ kg$^{-1}$ m$^3$ s$^{-2}$; the uncertainty in $a$, which is the most important source of error, has been evaluated as
\eqi \delta a \leq   \delta
a|_\mathcal{R} + \delta a|_s + \delta a|_{x_{\rm A}} ,\eqf with
\begin{equation}\left\{\begin{array}{lll}
\delta a|_\mathcal{R}\leq  \left(\rp{cx_{\rm A}}{s}\right)\delta \mathcal{R} = 466,758\ {\rm m},\\\\
\delta a|_s\leq \left(1+\mathcal{R}\right)\left(\rp{cx_{\rm
A}}{s^2 }\right)\delta s = 342,879\ {\rm m},\\\\
\delta a|_{x_{\rm A}} \leq \left(1+\mathcal{R}\right)\left(\rp{c}{s}\right)\delta x_{\rm
A}=621 \ {\rm m}.

 \lb{erra}
\end{array}\right.\end{equation}
Thus, \eqi \delta a \leq 810,259\ {\rm m}.\lb{longo}\eqf It is interesting to note that $\mathcal{R}$ and $s$
have a major impact on the overall uncertainty in $a$; our estimate has to be considered as conservative because
we adopted for $\delta s$ the largest value quoted in Ref.~\refcite{Kra06}.

The following considerations about the treatment of errors are in order.
\begin{itemize}
  \item Realistic upper bounds of the errors have been obtained by purposely linearly summing the various sources of uncertainty because of the existing correlations among the estimated parameters of the \psr system\cite{Kra06}
  \item The uncertainties of Table \ref{tavola} and used
throughout the paper are twice the parameter uncertainties given
by the software used in Ref.~\refcite{Kra06}; the authors of such a  work
believe that the real measurement uncertainties are actually
somewhat smaller than those quoted. If so, also our result for
$\Delta P$ would, in fact, be more accurate. By the way, in view of the
continuous timing of the \psr system it is likely
that in the near future the precision reached in determining its
parameters will allow to better constrain $\Delta P$, although it is difficult to quantify the extent of such an improvement because of the impact of $x_{\rm B}$ via $\mathcal{R}$.
\end{itemize}

\section{Applications}
In this Section we will look for deviations from the third Kepler law induced by small radial accelerations $A_r$, in general non-constant, induced by some modified models of gravity ($f(R)$, Yukawa-like fifth force, MOND). The effects of such accelerations, which are to be retained quite smaller than the Newtonian monopole, should, in principle, be treated perturbatively with, e.g., the Gauss perturbing equations for the variations of the Keplerian orbital elements of a test body. Indeed, while the average Newtonian acceleration for the \psr\ system amounts to
\eqi \left\langle A_{\rm N}\right\rangle=\rp{GM}{a^2\sqrt{1-e^2}}=446\ {\rm m\ s}^{-2},\eqf    the $f(R)$ acceleration (see Section \ref{fR}) is about three orders of magnitude smaller, the Yukawa-type term (see Section \ref{Yuki}) is five orders of magnitude smaller, while the MONDian correction (see Section \ref{mondi}) is even much smaller.

However, in view of the likely broad spectrum of readers, many of whom may not be  fully acquainted with the rather specialistic tools of celestial mechanics, we will simply work with \eqi {\dot\phi}^2 r = \rp{GM}{r^2} + A_r,\lb{perty}\eqf where $\phi$ is an azimuth angle, by neglecting the orbital eccentricity since the small corrections due to it would not affect the constraints set by the main terms.
\subsection{Extended theories of gravity: ``$f(R)$''}\lb{fR}
It has been suggested that cosmic speed-up can be  explained by generalizing general relativity  introducing in
the gravitational action terms non linear in the scalar curvature $R$ (see Refs.~ \refcite{c6,odintsov06} and references
therein). These theories are the so called $f(R)$ theories of gravity (or ``extended theories of gravity''),
where the gravitational Lagrangian depends on an arbitrary analytic function $f$ of the scalar curvature $R$.
If, on the one hand, extended theories of gravity seem to be appealing in cosmology, on the other hand, there is
the need of checking $f(R)$ predictions with Solar System tests which general relativity passed with flying
colors\cite{will}. Actually, the debate on the compatibility of these theories with Solar System tests is still
open (see Ref.~\refcite{faraoni06a} and Refs.~\refcite{ruggiero06,allemandi07} and references therein). Here we do not enter the
debate, but simply suggest how, on the basis of  the approach already outlined in Refs.~\refcite{ruggiero06,allemandi07},
the data from the double pulsar can be used, at least in principle, to constrain the allowable analytical forms
of the $f(R)$. We work in the Palatini formalism (for an approach to extended theories of gravity in the metric
formalism see, for instance, Ref.~\refcite{capozziello07} and references therein).
The field equations of $f(R)$ theories, in the Palatini formalism, which have the explicit form
\begin{eqnarray}
f^{\prime }(R) R_{(\mu\nu)}(\Gamma)-\frac{1}{2} f(R)  g_{\mu \nu
}&=&\kappa T_{\mu \nu }^{mat}  \label{ffv1}\\
\nabla _{\alpha }^{\Gamma }[ \sqrt{g} f^\prime (R) g^{\mu \nu })&=&0
\label{ffv2}
\end{eqnarray}
admit the spherically symmetric vacuum exact solution:
\begin{align}
ds^2=&-\left(1-\frac{2GM}{c^2r}+\frac{k r^2}{3}\right)c^2dt^2+\frac{dr^2}{\left(1-\frac{2GM}{c^2r}+\frac{k
r^2}{3}\right)} \notag \\ & +r^2d\vartheta^2+r^2\sin^2 \vartheta d\varphi^2, \label{eq:metrica1}
\end{align}
which is referred  to as Schwarzschild-de Sitter metric (see Ref.~\refcite{allemandi05}  for  details on the derivation of this solution). The parameters appearing in \rfr{eq:metrica1} are the mass $M$ of the spherically symmetric source of the gravitational field and $k$, which is related to the solutions $R=c_i$ of
the structural equation
\begin{equation}
f^{\prime }(R) R-2f(R)= 0. \label{eq:struct1}
\end{equation}
controlling the solutions of  \rfr{ffv1}. Namely, it is $k=c_i/4=R/4$.  Indeed, we may say that $k$ is
a measure of the non-linearity of the theory (if $f(R)=R$, \rfr{eq:struct1} has the solution $R=0
\rightarrow k=0$) In general relativity, the Schwarzschild-de Sitter solution corresponds to a spherically
symmetric solution of Einstein equations, with a "cosmological" term $\Lambda g_{\mu\nu}$, $\Lambda$ being the
cosmological constant. In practice, it is $\Lambda=-k$ in our notation. As we can see from \rfr{eq:metrica1},
the modifications to the solutions of the field equations due to $f(R)$ theories are given by a term
proportional to the Ricci scalar. In what follows, we write the gravitational potential in the form
\eqi U\equiv U_0+U_{\kappa}, \label{eq:potkappa1}\eqf with
\begin{equation}\left\{\begin{array}{lll}
U_0=-\rp{GM}{r},\\\\
U_{\kappa}=-\kappa r^2 \lb{poti}
\end{array}\right.\end{equation}
where $\kappa=k/3$, and we consider  $\Delta U \equiv U_{\kappa}$ as a perturbation. In doing so, we neglect the
effect of spatial curvature. From the perturbing potential $U_k$ in \rfr{poti} we obtain an entirely radial
perturbing acceleration \eqi {\boldsymbol{A}}=-2\kappa \boldsymbol{r}.\lb{ADM}\eqf
We may write
\eqi P_{\rm b} \equiv P^{(0)}+P^{(k)} \eqf
where the contribution $\Delta P \equiv P^{(k)}$, due to the non linearity of the Lagrangian,  by applying \rfr{perty}, turns out to be
\eqi \Delta P = -\rp{2\pi \kappa}{ {n^{(0)}}^3}, \label{eq:defp0fr1}\eqf from which $\kappa$ can be evaluated
\eqi \kappa =-\rp{\Delta P {n^{(0)}}^3}{2\pi} \label{eq:kappafr1} \eqf
Consequently, we may give the following estimates
\eqi \delta \kappa\leq \rp{ {n^{(0)}}^3 }{2\pi}\delta(\Delta P) + \rp{3|\Delta P| {n^{(0)}}^2}{2\pi}\delta n^{(0)}, \label{eq:deltakappafr1} \eqf
and
\eqi\kappa =(1\pm 7)\times 10^{-10}\ {\rm s}^{-2}  \label{eq:kappafr2}\eqf
Equivalently, restoring physical units
\eqi\kappa =(0.1\pm 0.8)\times 10^{-26}\ {\rm m}^{-2},  \label{eq:kappafr22}\eqf
or, rephrased in terms of the cosmological constant,
\eqi|\Lambda| = {3\kappa}=(0.3\pm 2.5)\times 10^{-26}\ {\rm m}^{-2}\lb{erlam} \eqf
The estimates of \rfr{erlam} can be used to set constraints on the analytical form of the $f(R)$, according to an approach described in Refs.~\refcite{ruggiero06,allemandi07}. For instance, if we consider the  Lagrangian
\begin{equation}
f(R)=R-\frac{\mu^4}{R}. \label{eq:frcarrol1}
\end{equation}
which mimics cosmic acceleration (even if it has well known instabilities  see, e.g., Ref.~\refcite{faraoni06b}), thanks to the estimates on $\kappa$ we may set a limit on the parameter $\mu$. We get
\begin{equation}
|\mu| \leq 10 ^ {-20} \ {\rm eV}. \label{eq:carrol4}
\end{equation}
We remark that the estimates  of \rfr{erlam} are of the same order of magnitudes of those obtained in Refs.~\refcite{ruggiero06,allemandi07}, where data coming from Solar System observations were used. So, as we did there, we can say that also data coming from pulsars are hardly fit to constrain $f(R)$, since the value \rfr{eq:carrol4} is remarkably greater than estimate\cite{c6} $\mu \simeq
10 ^ {-33} \ {\rm eV}$, needed for $f(R)$ gravity to explain the acceleration of the Universe without requiring dark matter.

\subsection{Yukawa-like fifth force}\label{Yuki}
Modifications of the gravitational potential leading to a Yukawa-like term arise in different contexts, such as
brane-world models, scalat-vector-tensor theories of gravity, string theory, study of cosmological defects (see,
e.g., Refs.~\refcite{adelberger07,sereno06} and references therein). In general, a Yukawa-like force can be
obtained from the potential\cite{Ber06}
\eqi U\equiv U_0+U_{Y}, \label{eq:potyuk1}\eqf
with
\begin{equation}\left\{\begin{array}{lll}
U_0=-\rp{GM}{r},\\\\
U_{Y}=-\rp{GM}{r}\left[\alpha \exp\left(-\frac{r}{\lambda}\right)\right],\lb{eq:yuk2}
\end{array}\right.\end{equation}
where $|\alpha|$ is the strength of the Yukawa-like force and $\lambda$ its range\cite{adelberger}. The
parameters $\alpha, \lambda$ have been bounded thanks to many laboratory experiments and observations, at
different scales (see Ref.~\refcite{bertolami03} and references therein). Here we show how the strength $|\alpha|$ of
the Yukawa-like force can be constrained at a scale $\lambda \approx a \approx 10 ^9 \rm{m}$, which is the
length scale of the double pulsar \psr.
The perturbing potential $\Delta U \equiv U_{Y}$ in \rfr{eq:yuk2} leads to the following expression for the
radial component of the corresponding perturbing acceleration: \eqi A_{\rm
Y}=-\rp{GM\alpha}{r^2}\left(1+\rp{r}{\lambda}\right)\yu.\lb{yacc}\eqf
From \rfr{yacc} it is possible to obtain, in the circular orbit approximation, \eqi P = \pkp\left[1-\rp\alpha
2\left(1+\rp{r}{\lambda}\right)\yu\right]. \label{eq:yuk22}\eqf
Hence, on writing  \eqi P_{\rm b} \equiv P^{(0)}+P^{(Y)}, \label{eq:yuk3} \eqf from \rfr{perty}
we obtain for the contribution $\Delta P \equiv P^{(Y)}$ the following expression
\eqi \Delta P = -\rp{\pi\alpha}{n^{(0)}}\left(1+\rp{r}{\lambda}\right)\yu.\eqf By assuming $\lambda\approx r \approx
a$, we get
\begin{equation}\left\{\begin{array}{lll}
\alpha \approx -\rp{\Delta P n^{(0)} \exp(-1)}{2\pi}=0.7\times 10^{-4},\\\\
\delta\alpha\leq \rp{n^{(0)}\exp(-1)}{2\pi}\delta(\Delta P) + \rp{|\Delta P| \exp(-1)}{2\pi}\delta n^{(0)}=5.5\times 10^{-4}.
 \lb{yukaz}
\end{array}\right.\end{equation}
These estimates for $\alpha$, should be compared with those obtained at different scales\cite{bertoloami05,rey05,Ior07}.
In particular, in Ref.~\refcite{rey05} it is showed that $|\alpha| \leq 10^{-9}$ at a
scale corresponding to that of the double pulsar. In other words, our results are not compelling, since they are
orders of magnitude greater than the best estimates already available. This means that the length scale of the
hypothesized Yukawa-like fifth force is much greater or much smaller  than the length-scale of the  double
pulsar \psr\ ($a=0.006$ AU).  The results obtained in  Ref.~\refcite{Ior07} for the Solar System point towards the second hypothesis, i.e. $\lambda < 0.006$ AU.

\subsection{MOND}\label{mondi}
MOND predicts that the gravitational acceleration $\bds A_g$ felt by a particle in the field of a distribution of mass is
\eqi\bds A_g=\rp{\bds A_{\rm N}}{\mu\left(\rp{A_g}{A_0}\right)},\lb{mondacc}\eqf where
$\bds A_{\rm N}$ is the Newtonian acceleration, $A_0$ is an acceleration scale which different, independent ensembles of observations set to\cite{San02} $A_0=1.2\times 10^{-10}$ m s$^{-2}$, and $\mu(x)$ is an interpolating function which approximates 1 for $x\gg 1$, i.e. for accelerations larger than $A_0$; for $x\ll 1$ $\mu(x)=x$, so that in such a strongly MONDian regime $A_g\approx \sqrt{A_{\rm N}A_0}$.
For a quite general class of interpolating functions, $\mu(x)$
can be cast into the form\cite{Mil83}
\eqi\mu(x)\approx 1-k_0\left(\rp{1}{x}\right)^q,\lb{milgrom}\eqf
which yields a modified gravitational acceleration\cite{Tal88}
\eqi \bds A_g\approx \bds A_{\rm N}\left[1+k_0\left(\rp{A_0}{A_{\rm N}}\right)^q\right].\lb{accmilgrom}\eqf
Note that the most commonly used expressions for $\mu(x)$, i.e.\cite{Mil83}
\eqi\mu(x)=\rp{x}{\sqrt{1+x^2}},\lb{Milmu}\eqf and\cite{Fam05}
\eqi\mu(x)=\rp{x}{1+x},\lb{Fammu}\eqf can be obtained from \rfr{accmilgrom} for $k_0=1/2, q=2$ and
$k_0=1, q=1$, respectively. A useful review dealing, among other things, with many attempts to theoretically justify MOND is\cite{Bek06}.
Again, we may write \eqi P_{\rm b} \equiv P^{(0)}+P^{(\rm M)}, \label{eq:mond11} \eqf and, from \rfr{perty} with $A_r$ given by \rfr{accmilgrom}, on
setting  $e\rightarrow 0$, it is straightforward to obtain for the perturbation $\Delta P \equiv P^{(M)}$, \eqi
\Delta P = -\rp{\pkp k_0}{2}\left(\rp{A_0}{A_{\rm N}}\right)^q.\eqf By defining \eqi H_{{k_0},q}\equiv -\rp{\pkp
k_0}{2}\left(\rp{A_0}{A_{\rm N}}\right)^q,\eqf it turns out that both \rfr{Milmu} and \rfr{Fammu} are compatible
with the \psr\ data: indeed,
\begin{equation}\left\{\begin{array}{lll}
H_{{1/2},2}=(-1.610\pm 0.008)\times 10^{-22}\ {\rm s},\\\\
H_{1,1}=(-1.431\pm 0.004)\times 10^{-19}\ {\rm s}.
\lb{cazzorotto}
\end{array}\right.\end{equation}
By the way, from \rfr{cazzorotto} it is clear that the \psr\ system is definitely unsuitable for testing MOND because
discrepancies as small as those of \rfr{cazzorotto} will never be measurable, falling well below even the precision in determining $P_{\rm b}$.  The same holds also for other binary systems, given the obtainable accuracy in determining their orbital periods.
Solar System dynamics\ct{sereno06} has proven to be more effective in constraining the interpolating function for $x>>1$.

\section{Discussion and conclusions}
In this paper we looked for deviations from the third Kepler law in the \psr double pulsar system finding a
discrepancy $\Delta P = -1.772341\pm 13.153788$ s between the phenomenologically determined orbital period
$P_{\rm b}$ and the purely Keplerian one $\pkp$. While $P_{\rm b}$ is determined to a $10^{-6}$ s level, $\pkp$\
can be considered known only at $\approx 10^1$ s level after a conservative propagation of the uncertainties in
the system's parameters entering its expression. The consistency of our analysis is assured by the fact that the
values of the orbital parameters involved in the calculation of $\pkp$ have been determined independently of the
third Kepler law itself. The major source of error in $\pkp$\ is due to the semimajor axis which, in turn, is
mainly affected by the ratio $\mathcal{R}$ of the pulsars' masses, which, in turn, depends on the poorly
determined projected semimajor axis of B, and by $\sin i$. Continuous timing of the \psr system might reduce
such errors in the near future, perhaps yielding more precise results.

We used $\Delta P$ to constrain various theories of modified gravity like $f(R)$, Yukawa-type interactions and MOND. The bound on the parameter $\kappa$, that in $f(R)$ framework is a measure of the deviation
of the theory from general relativity, is $|\kappa|\leq 0.8\times 10^{-26}$ m$^{-2}$; differently speaking, the
bound on $\kappa$ can be seen as a bound on the cosmological constant $|\Lambda|=3 \kappa$. By assuming a range
$\lambda$ of the order of the \psr system's size, we get $|\alpha|\leq 5\times 10^{-4}$ for the strength of a
putative Yukawa-like fifth force.  In regard to MOND, the \psr\ system is neatly unsuitable to test it because
the order of magnitude of the predicted MONDian discrepancies is $\Delta P_{\rm MOND}\approx 10^{-19}-10^{-22}$
s.

\section*{Acknowledgements}
L.I. gratefully thanks J. Katz for stimulating discussion. M.L.R  acknowledges financial support from the
Italian Ministry of University and Research (MIUR) under the national program 'Cofin 2005' - \textit{La pulsar
doppia e oltre: verso una nuova era della ricerca sulle pulsar}.  Both L.I. and M.L.R. gratefully acknowledge the referee for his/her useful
comments which notably improved the manuscript.


\end{document}